\documentclass[aps,prd,twocolumn,showpacs,nofootinbib,amsmath,amssymb,floatfix,superscriptaddress,showkeys]{revtex4}
\usepackage{graphicx}% Include figure files
\usepackage{dcolumn}% Align table columns on decimal point
\usepackage{bm}% bold math
\usepackage{captcont}
\usepackage{xcolor}
\usepackage{threeparttable}

\usepackage{epstopdf}
\usepackage{mathtools}
\usepackage{natbib}
\usepackage{ulem}
\usepackage{color,txfonts}
\definecolor{blue0}{rgb}{0,0,0.6}
\usepackage[colorlinks,linkcolor=blue0,anchorcolor=blue0,citecolor=blue0,urlcolor=blue0]{hyperref}

\newcommand{\jcap}{J. Cosmol. Astropart. Phys.}

\newcommand{\mnras}{Mon. Not. R. Astron. Soc.}

\newcommand{\aap}{Astron. Astrophys}

\newcommand{\nar} {NewA Rev.}

\newcommand{\hess}{H.E.S.S. }

\begin{document}
\title{Constraints on axion-like particle properties with very high energy gamma-ray observations of Galactic sources}

\author{Yun-Feng Liang}
\affiliation{Key Laboratory of Dark Matter and Space Astronomy, Purple Mountain Observatory, Chinese Academy of Sciences, Nanjing 210008, China}
\author{Cun Zhang}
\affiliation{Key Laboratory of Dark Matter and Space Astronomy, Purple Mountain Observatory, Chinese Academy of Sciences, Nanjing 210008, China}
\affiliation{School of Physics, Nanjing University, Nanjing, 210092, China}
\author{Zi-Qing Xia}
\email{xiazq@pmo.ac.cn}
\affiliation{Key Laboratory of Dark Matter and Space Astronomy, Purple Mountain Observatory, Chinese Academy of Sciences, Nanjing 210008, China}
\affiliation{School of Astronomy and Space Science, University of Science and Technology of China, Hefei, Anhui 230026, China}
\author{Lei Feng}
\affiliation{Key Laboratory of Dark Matter and Space Astronomy, Purple Mountain Observatory, Chinese Academy of Sciences, Nanjing 210008, China}
\author{Qiang Yuan}
\affiliation{Key Laboratory of Dark Matter and Space Astronomy, Purple Mountain Observatory, Chinese Academy of Sciences, Nanjing 210008, China}
\affiliation{School of Astronomy and Space Science, University of Science and Technology of China, Hefei, Anhui 230026, China}
\author{Yi-Zhong Fan}
\email{yzfan@pmo.ac.cn}
\affiliation{Key Laboratory of Dark Matter and Space Astronomy, Purple Mountain Observatory, Chinese Academy of Sciences, Nanjing 210008, China}
\affiliation{School of Astronomy and Space Science, University of Science and Technology of China, Hefei, Anhui 230026, China}

\date{\today}

\begin{abstract}
Axion-like particles (ALPs) can oscillate to photons and vice versa in electromagnetic fields.
The photon-ALP oscillation provides an attractive solution to the apparent transparency of the Universe to TeV photons.
The allowed parameter regions for the ALP mass $m_{\rm a}\leq 10^{-7}$ eV have been tightly constrained by the Fermi-LAT and H.E.S.S observations of some extragalactic sources.
In this work we show for the first time that the H.E.S.S observations of some TeV sources in the Galactic plane exclude  the highest ALP mass region (i.e., $m_{\rm a}\sim {\rm a ~few\times 10^{-7}}$ eV) that accounts for the TeV transparency of the Universe. The upcoming CTA observations of the Galactic TeV sources are shown to be able to improve the constraints significantly.
\end{abstract}

\maketitle

\section{Introduction}
\label{sec1}

The axion, a type of pseudoscalar particle beyond the standard model (SM), is postulated to solve a puzzle in quantum chromodynamics (QCD) known as ``the strong CP problem" \cite{pq1977, Weinberg1978, pq2008}.
If the critical axion mass is about ${10^{-11}}$ times the electron mass, where there is no lighter particle to decay into, it is predicted that the Universe would be abounded with the cold Bose Einstein condensate of primordial axions. Consequently, axions could plausibly account for all or a significant fraction of the cold dark matter (DM) \cite{DMA,DMA2012}.
More generalized  axionlike-particles (ALPs), which have similar properties with axions, are predicted in several string-theory-motivated extensions of the SM as an alternative to axions to solve the DM problem \cite{alp1982, alp2012, alp2014, alp2017}.
Axions and ALPs have an interesting property that they can oscillate into photons and vice versa in electromagnetic fields via the Primakoff process \cite{axionf}. The Lagrangian of the interaction can be written as
$\mathcal{L}=g_{a\gamma}\vec{E}\cdot\vec{B}{a}$,
where $\vec{E}$, $\vec{B}$ and $a$ are electric, magnetic and axion (or axion-like) fields, and $g_{a\gamma}$ is the coupling constant.
While the axion mass $m_{a}$ is proportional to the coupling constant $g_{a\gamma}$, these two parameters are not necessarily related to each other for ALPs.

Great efforts have been made to probe axions and ALPs via the Primakoff effect.
Some laboratory experiments, such as ALPS, CAST and ADMX, have been carried out to detect axions and ALPs \cite{expph2013,exp2015,light2011,ALPSI2010,CAST2017,ADMX2010}.
Astrophysical observation data are also used to probe axions/ALPs \cite{belikov11alp,hess13pks2155,reesman14alp,payez15_1987a,berenji16alp_ns,fermi16alp,meyer17sne,zc16alp,Majumdar17psr2,ALP2018,ic443alp}.
However, no positive signal has been identified so far and upper limits have been reported.

Besides being DM candidates, the ALPs also provides a possible solution to the apparent transparency of the Universe to TeV photons \cite{mirizzi07alp_tev,angelis07,angelis09,sc09,Kohri:2017ljt}. 
TeV photons may interact with extragalactic background light (EBL) in their routes to the Earth, thus the Universe is opaque to sufficiently distant TeV sources. The TeV transparency is dependent on the photon energy. However, some indications showed that the opacity of the Universe is lower than that expected by the EBL model \cite{magic3c279}.
This anomalous transparency can be easily explained if existing ALPs \cite{mirizzi07alp_tev,angelis07,angelis09,sc09,Kohri:2017ljt}.
TeV photons can oscillate into ALPs and vice versa in the presence of cosmic magnetic fields. The ALPs produced near the source can unimpededly pass through long distances, then convert back to TeV photons before reaching the earth, thus lower the opacity of the universe.
The parameter space where ALP can account for the transparency has been calculated in the literature \cite{Meyer13}.
Both the aforementioned laboratory experiments and the $\gamma-$ray observations of some extragalactic sources have effectively narrowed down the parameter space. In particular, with the $\gamma$-ray spectrum of NGC 1275, the Fermi-LAT collaboration has excluded the coupling $g_{a\gamma}$ above $5 \times 10^{-11}\,{\rm GeV}^{-1}$ for ALP masses $0.5 - 5$ neV \cite{fermi16alp} (see also \cite{zc16alp} for the additional constraint set by the Fermi-LAT observation of
PKS 2155-304).
Moreover, the H.E.S.S. observation of PKS 2155-304 has yielded an upper limit of the coupling $g_{\gamma{a}}<2.1\times10^{-11}\,{\rm GeV^{-1}}$ for $m_{\rm a}\sim 15 - 60\,{\rm neV}$ \cite{hess13pks2155}. Nevertheless, the high end part of the allowed parameter space (i.e., $m_{\rm a}>10^{-7}$ eV) for the TeV transparency has not been independently probed by the gamma-ray data, yet. To achieve a constraint better than CAST (i.e., $g_{a\gamma}<6\times 10^{-11}~{\rm GeV}^{-1}$), we need to observe the photons at the energies of $\sim 2~{\rm TeV}~(m_{\rm a}/10^{-7}~{\rm eV})^{2}(g_{a\gamma}/5\times 10^{-11}~{\rm GeV^{-1}})^{-1}(B_{\rm T}/1\mu G)^{-1}$, where $B_{\rm T}$ is the strength of the large scale magnetic filed. 
Therefore the H.E.S.S observations of some TeV sources in the Galactic plane are ideal probe of the photon-ALP oscillation effect for $m_{\rm a}\sim 10^{-7}$ eV. Hence in this work we focus on the bright H.E.S.S sources in the Galactic plane and search for the potential irregularities caused by the photon-ALP oscillation in the gamma-ray spectra.

\section{Photon-ALP Oscillation in The Milky Way Magnetic Field}
\label{sec:sec2}
ALPs have similar property to axions: they can convert into photons (and vice versa) in the external magnetic field.
For a coherent magnetic field with size $l$ and a transversal strength $B_{\rm T}$, the survival probability that a photon is measured for an initially polarized incoming photon with energy $E_\gamma$ is approximated as \cite{axionf,axionf1}
\begin{eqnarray}\label{eq:p}
P_{\rm ALP}&=&1-P_{\gamma \rightarrow a} \\ \nonumber
&=&1-\frac{1}{1+{E_{\rm c}^2}/{E_\gamma^2}} \sin^2 \left[\frac{g_{a\gamma} B_{\rm T} l}{2}  \sqrt{1+\frac{E_{\rm c}^2}{E_\gamma^2}} \right].
\end{eqnarray}
The oscillation behavior becomes most significant near the characteristic energy $E_{\rm c}$ which is defined as
\begin{eqnarray}
E_{\rm c}=\frac{\left| m_a^2-w_{\rm pl}^2\right|}{2g_{a\gamma} B_{\rm T}},
\end{eqnarray}
where $w_{\rm pl}^2=4\pi\alpha n_{\rm e}/m_{\rm e}$ is the plasma frequency.
The Milky Way magnetic fields are of the order of $B_{\rm T} \sim 1\,\mu{\rm G}$, with the thermal electron density $n_{\rm e}=0.1\,{\rm cm^{-3}}$ and the current upper limit $g_{a\gamma}\sim10^{-10}\,{\rm GeV}^{-1}$, a characteristic energy of $E_{\rm c}=1~{\rm TeV}$ corresponds to a ALP mass of $m_a=10^{-7}\,{\rm eV}$.
Thus the \hess is most sensitive to the ALPs around this mass.
In our analysis, rather than using the Eq. (\ref{eq:p}), we explicitly solve the evolution equation for photon-ALP beam as in Ref. \cite{axionf,axionf1} to calculate the survival probability.
For simplicity, the source photons are assumed to be unpolarized. Such an assumption would yield conservative results.

The Galactic magnetic field consists of a large-scale regular component and a small-scale random component.
In our work, we neglect the later since its coherence length is much smaller than the photon-ALP oscillation length.
For the regular magnetic field, we take into account the model developed by Jansson \& Farrar \cite{Bfield1} that has been widely adopted in similar researches \cite{fermi16alp,meyer17sne,ALP2018}.

\section{samples \& methods}
\label{sec2}

%*******************table.1****************************
\begin{table*}[!t]
\caption{Selected gamma-ray sources used for our analysis.}
\begin{tabular}{ccccccccccccc}
\hline
\hline
Name     &  $l$  &   $b$  &  Flux  &  $d^a$ &  Spectrum$^b$   &   $\alpha$  &  $E_0$   &   $\beta$ &  $E_{\rm cut}$   & Type$^c$  & Reference \\
       &   [$^\circ$] & [$^\circ$] &  [${\rm ph/cm^2/s}$]  & [kpc] &  &  &  [TeV]   &    &  [TeV]   &   &  \\
\hline

HESS J1713-397  & 347.28 & -0.38  & 0.66 &  1.0  &  ECPL  &  2.17  &  1.0     &  2.0    & 16.5  & SNR    & \cite{1713} \\
HESS J1826-148  & 16.88  & -1.28  & 0.03 &  2.5  &  ECPL  &  2.06  &  1.0     &  1.0    & 13    & HMXRB  & \cite{1826} \\
HESS J1731-347  & 353.57 & -0.62  & 0.16 &  3.2  &  PL    &  2.32  &  0.783   &  $-$    & $-$   & SNR    & \cite{1731} \\
HESS J1825-137  & 17.82  & -0.74  & 0.17 &  3.9  &  ECPL  &  2.26  &  1.0     &  1.0    & 24.8  & PWN    & \cite{1825} \\
HESS J1813-178  & 12.81  & -0.03  & 0.06 &  4.7  &  PL    &  2.09  &  1.0     &  $-$    & $-$   & PWN    & \cite{18041337} \\
HESS J1514-591  & 320.33 & -1.19  & 0.15 &  5.2  &  PL    &  2.27  &  1.0     &  $-$    & $-$   & PWN    & \cite{1514} \\
HESS J1804-216  & 8.40   & -0.03  & 0.25 &  6.0  &  PL    &  2.72  &  1.0     &  $-$    & $-$   & UNID   & \cite{18041337} \\
HESS J1303-631  & 304.24 & -0.36  & 0.17 &  6.6  &  ECPL  &  1.5   &  1.0     &  1.0    & 7.7   & PWN    & \cite{1303} \\
HESS J1837-069  & 25.18  & -0.12  & 0.13 &  6.6  &  PL    &  2.27  &  1.0     &  $-$    & $-$   & PWN    & \cite{18041337} \\
HESS J1640-465  & 338.32 & -0.02  & 0.09 &  8.6  &  ECPL  &  2.11  &  1.0     &  1.0    & 6     & SNR    & \cite{1640} \\
\hline
\end{tabular}

\begin{tablenotes}
\item $^{\rm a}$ The distances $d$ are extracted from the website \url{http://tevcat.uchicago.edu/}.
\item $^{\rm b}$ The spectral types and parameters ($\alpha$, $E_0$, $\beta$ and $E_{\rm cut}$) are from the literature of the last column.
\item $^{\rm c}$ SNR: supernova remnant; HMXRB: high-mass X-ray binary; PWN: pulsar wind nebula; UNID: unidentified.
\end{tablenotes}

\label{tab:10srcs}
\end{table*}
%################################################################################

\begin{figure*}[!t]
\centering
\includegraphics[width=0.45\textwidth]{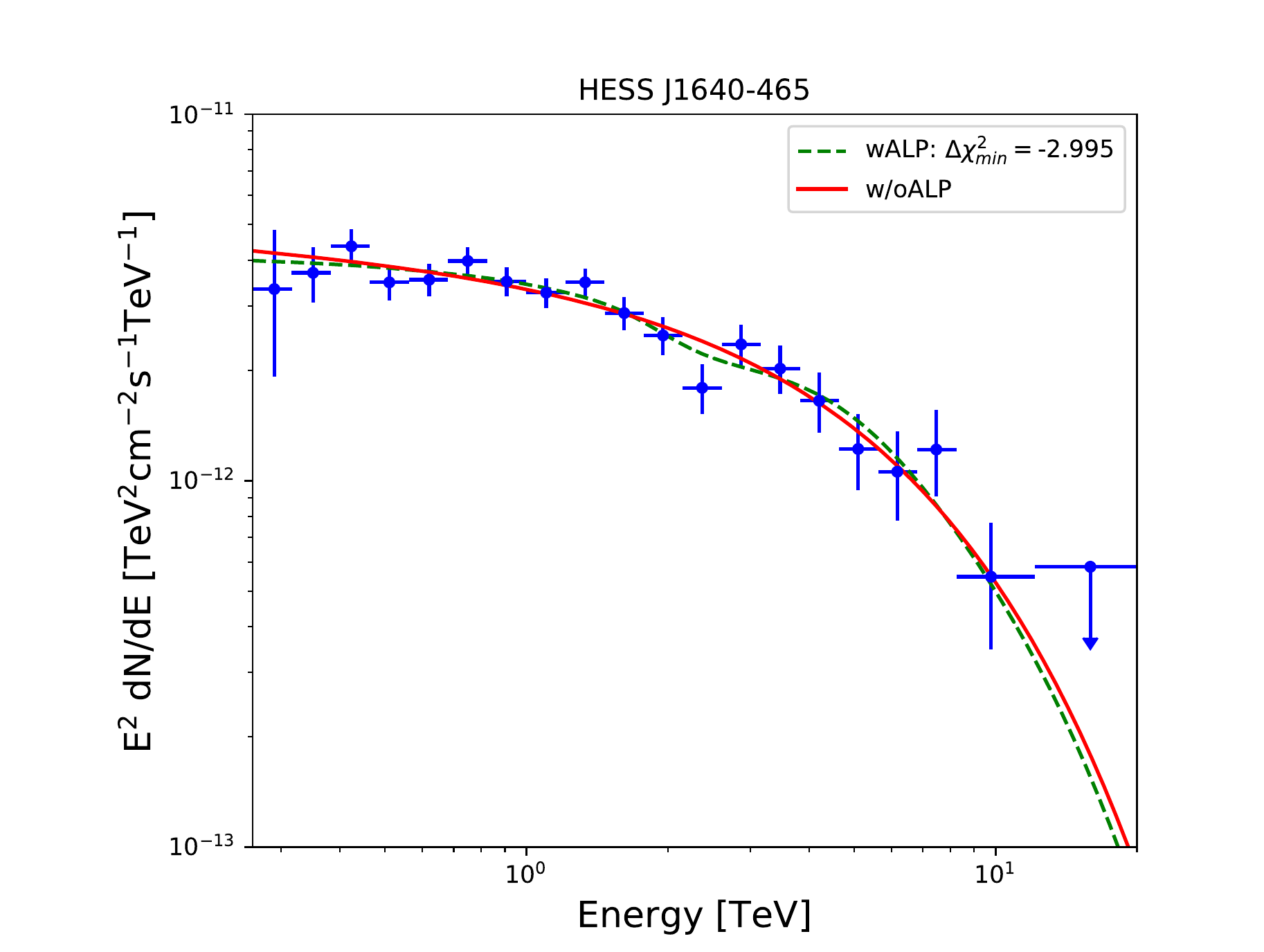}
\includegraphics[width=0.45\textwidth]{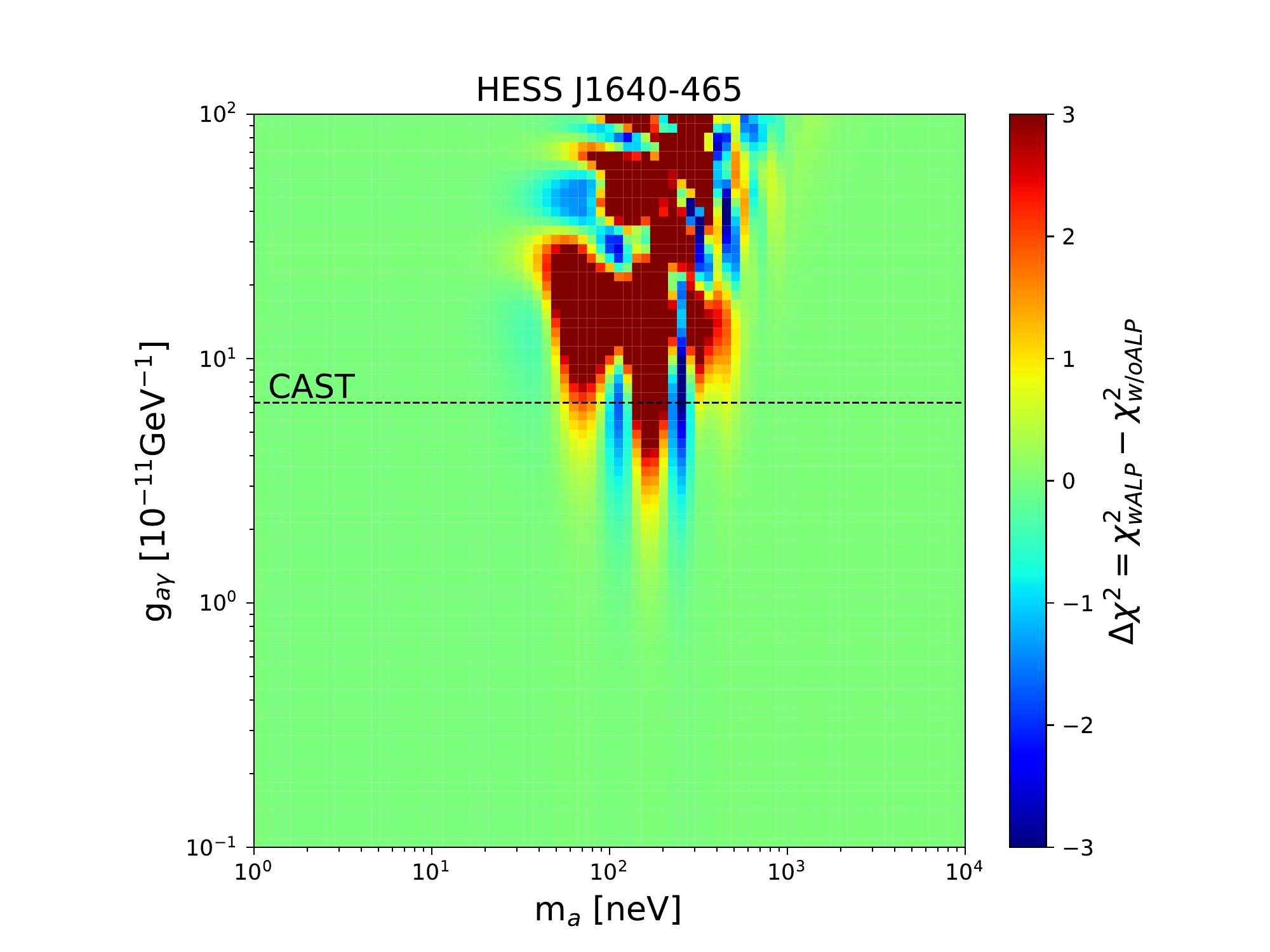}
\caption{The H.E.S.S. spectral energy distributions ({\it left panel}) and the map of $\Delta\chi^2$ as a function of ALP mass $m_{a}$ and photon-ALP coupling constant $g_{a\gamma}$ ({\it right panel}) for HESS J1640-465. In the right panel, the upper limit of $g_{a\gamma}$ set by CAST \cite{CAST2017} experiment is also plotted as a reference (black dashed line).}
\label{fig:1640}
\end{figure*}

High Energy Stereoscopic System (H.E.S.S.) is a system of Imaging Atmospheric Cherenkov Telescopes that observe gamma rays in the energy range from tens of GeV to tens of TeV \cite{hess}. In this work we focus on the \hess observation of bright Galactic sources, though some other devices, such as VERITAS \cite{2002APh....17..221W}, MAGIC \cite{2004NewAR..48..339L} and HAWC \cite{2013APh....50...26A}, are also sensitive to the TeV $\gamma$-rays. We collect the \hess observation spectra from the literature\footnote{All the spectrum data are extracted from {\tt Aux. information and data points} on the website \url{https://www.mpi-hd.mpg.de/hfm/HESS/pages/publications/pubs_jour.shtml}}. The oscillation effect due to ALP-photon conversion is significant only when the $\gamma$-rays have traversed considerable magnetic fields. To increase the amplitude of the oscillation effect, we limit our sample to those with distances greater than 1 kpc.
In addition, to guarantee sufficient statistics, we select only the bright sources.
Totally 10 sources are considered in our work.
Most of them are supernova remnants (SNRs) or pulsar wind nebulae (PWNe).
The basic information of these sources are listed in Table \ref{tab:10srcs}.

To examine whether the oscillation signals exist, we fit each \hess spectrum with two types of models, the background model (the null model) and the ALP model (the alternative model). The background model is for the case that intrinsically no ALP exists and these TeV sources can be well fitted with smooth astrophysical spectra. The signal model is the superposition of the ALP effect on the background spectrum (i.e., the modulation on the spectrum).

In this work, the intrinsic spectrum is described by two types of functions, a {\tt power law} (PL)
\begin{equation}
\Phi(E)=N_0\left({E}/{E_{0}}\right)^{-\alpha},
\label{eq:pl}
\end{equation}
or an {\tt exponential cutoff power law} (ECPL),
\begin{equation}
\Phi(E)=N_0\left({E}/{E_{0}}\right)^{-\alpha}{\exp}(-(E/E_{\rm cut})^{\beta}),
\label{eq:epl}
\end{equation}
where $N_0$, $\alpha$, $\beta$ and $E_{\rm cut}$ are the free parameters in the fittings. For each source, the function type of the intrinsic spectrum is chosen as that in the literature. The spectral types and the corresponding parameter values reported therein are also presented in Table \ref{tab:10srcs}. For the ALP model, the spectrum is expressed as
\begin{equation}
{{\Phi}_{\rm wALP}(E)=P_{\rm ALP}(g_{a\gamma}, m_{a}, E)\cdot{\Phi}_{\rm w/oALP}}(E),
\label{eq:ALP}
\end{equation}
where the $P_{\rm ALP}(g_{a\gamma}, m_{a}, E)$ is the survival probability.

The energy dispersion of ground based Cherenkov telescope like \hess is usually large and thus should be considered in the ALP analysis.
We approximate the energy dispersion function to be a Gaussian with its $\sigma$ being the energy resolution of the instrument.
The \hess energy resolution reported in the literature range from 10\% to 20\% \cite{hess13pks2155,hess_line1,hess_line2}. To be conservative, we adopt a value of 20\% in our analysis.
The systematic uncertainty due to the choice of the energy resolution will be discussed in Appendix \ref{sec:uncertainties} by testing some other values of energy resolution.
After convolving the energy dispersion function $D(E',E)$, we get the final model expected spectrum
\begin{equation}
{\Phi'(E')}=D(E',E)\otimes{\Phi(E)}.
\label{eq:Deff}
\end{equation}

The $\chi^2$ fit is utilized to analyze the spectrum of each source. The $\chi^2$ value is defined as
\begin{equation}
\chi^2=\sum_i^{N_{\rm bins}}\frac{(\Phi'(E_i)-\tilde{\phi}_i)^2}{\delta_i^2},
\label{eq:chi2}
\end{equation}
where $E_i$ is the geometrical central energy, $\tilde{\phi}_i$ and $\delta_i$ are the observed flux and its uncertainty in the $i$-th bin, respectively.
To validate the $\chi^2$ fit, we only consider the bins with $\tilde{\phi}_i>3\delta_i$ \footnote{According to Poisson statistics, this condition requires the photon number $N_i\gtrsim10$.}.
Smaller $\chi^2$ indicates the selected model can fit the observation better.
Thus whether the ALP model is favored or not can be determined through the quantity $\Delta\chi^2=\chi_{\rm wALP}^2-\chi_{\rm w/oALP}^2$.

For deriving the exclusion region of the ALP parameters, we calculate the difference $\lambda(m_a,g_{a\gamma})$ between the $\chi^2$ values for each set of ALP parameters ($m_a$, $g_{a\gamma}$) and the best fit over the whole parameter space.
If $\lambda>\lambda_{\rm thr}$ for a set of ($m_a$, $g_{a\gamma}$), it is considered as excluded at the 95\% confidence level (C.L.).
The threshold value is set to $\lambda_{\rm thr}=15.5$ in our analysis according to Monte-Carlo simulations (see Appendix \ref{sec:threshold} for details).

{}
\section{results}
\label{sec5}

With the procedure described above, we fit the spectrum of each source with two sets of spectral models. For the ALP model, we scan a grid of $m_\chi$ and $g_{a\gamma}$ and derive the $\Delta\chi^2$ for each set of parameters.
In the left panel of Figure \ref{fig:1640}, the spectrum of a representative source, HESS J1640-465, is shown, together with the best-fit background model (red line) and ALP model (green line).
The map of $\Delta\chi^2$ value as a function of ALP mass $m_{a}$ and photon-ALP coupling constant $g_{a\gamma}$ is demonstrated in the right panel of Figure \ref{fig:1640}.
A negative $\Delta\chi^2$ means that the ALP model results in a better fit to the observed spectrum than the background one. Meanwhile, a positive $\Delta\chi^2$ suggests that the fit with ALP effect is worsen, indicating such a hypothesis is disfavored.
As is shown, for part of the ALP parameters (blue region), the inclusion of the modulation effect does improve the goodness of fit. However, none of these sources shows a signal with $TS_{\rm max}>9$ (the test statistics, i.e. TS, is defined as $TS=-\Delta\chi^2$).
Thus, in the following, we focus on the exclusion region (roughly the red region in the $\Delta\chi^2$ map of Figure \ref{fig:1640}).
The horizontal line in Figure \ref{fig:1640} represents the upper limits of $g_{a\gamma}$ set by the CAST experiment \cite{CAST2017}.
Below this limit, the source HESS J1640-465 imposes almost the tightest constraints on ALP properties due to its longest distance among the sample and relatively high flux (see Appendix \ref{sec:9srcs} for $\Delta\chi^2$ maps of other 9 sources).

\begin{figure}[!t]
\centering
\includegraphics[width=0.95\columnwidth]{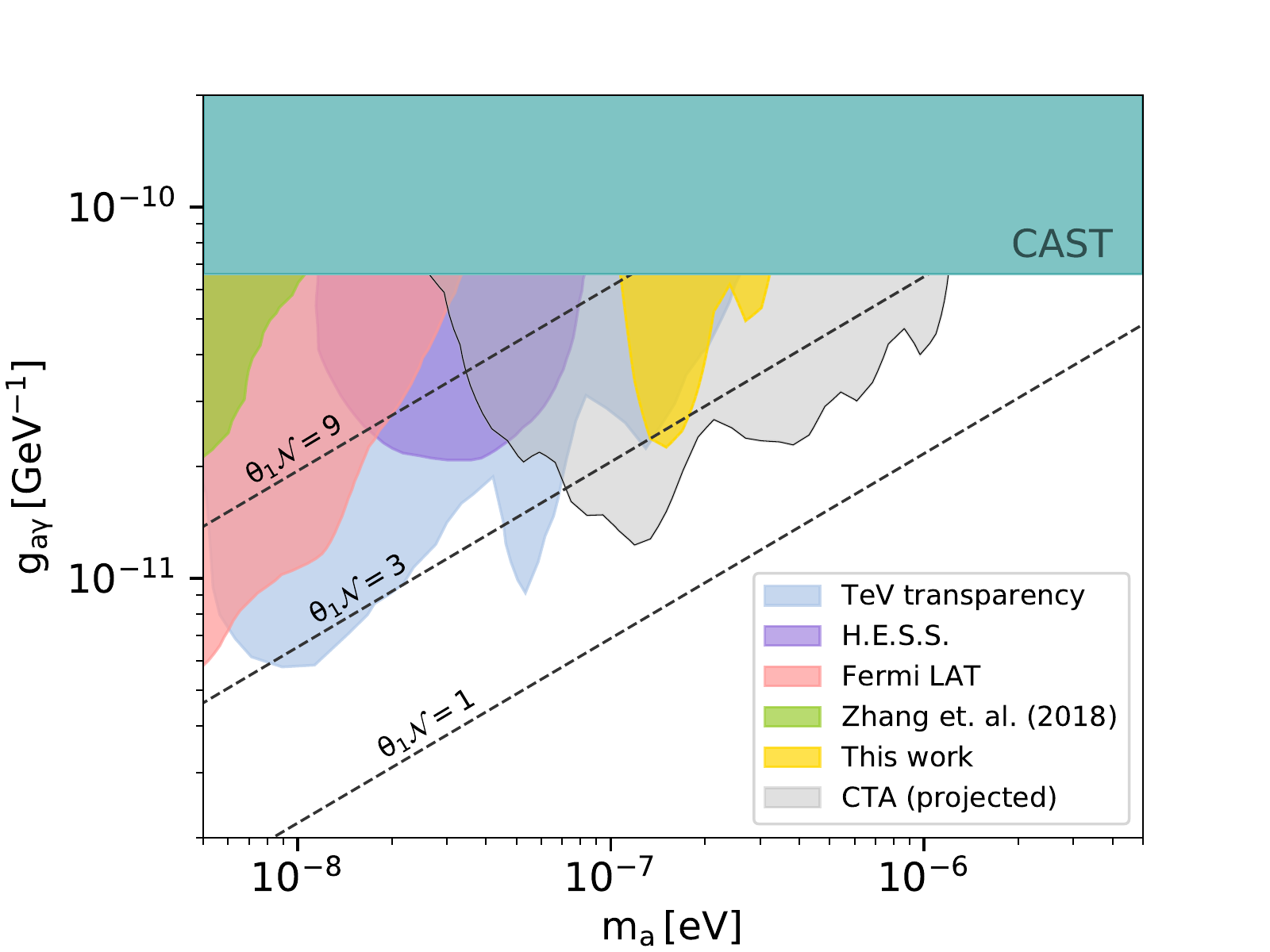}
\caption{Comparisons of the current limits on the ALP parameters. The red and green areas are obtained by Fermi-LAT observations
of NGC 1275 \cite{fermi16alp} and PKS 2155-304 \cite{zc16alp}, respectively. The H.E.S.S. observation of PKS 2155-304 leads to the exclusion region shown as purple \cite{hess13pks2155}.
The dashed line labeled $\theta_1\mathcal{N}=1$ ($\theta_1\mathcal{N}=3$, $\theta_1\mathcal{N}=9$) denotes the parameter space where the ALP dark matter accounts for all (1/9,1/81) of the dark matter. 
The light blue region is the parameter space where the low gamma-ray opacity of the universe can be explained by the ALPs \cite{Meyer13}. The result derived in this work is shown as yellow. In the plot, we also present the projected exclusion region assuming 50-hour observations of the 10 Galactic sources by future CTA mission.}
\label{fig:cpr}
\end{figure}

By summing the $\Delta\chi^2$ maps of the 10 sources, we derive combined constraints on the ALP parameters. The results are shown in Figure \ref{fig:cpr}. The yellow region is excluded by the combined analysis at 95\% confidence level.
In comparison with the previous constraints by other targets and instruments \cite{hess13pks2155,fermi16alp,zc16alp}, our results restrict the ALP properties at higher $m_{\rm a}$ and significantly improve the current limits around $m_{\rm a}\sim150\,{\rm neV}$.
Intriguingly, our exclusion region covers the high ALP mass space to interpret the TeV transparency (see the light blue region in Figure \ref{fig:cpr}).

The above results are based on the Galactic magnetic field model of Jansson \& Farrar \cite{Bfield1} that has been widely adopted in relevant investigations \cite{fermi16alp,meyer17sne,ALP2018}.
We have estimated the effect of the magnetic field models on the final constraints. With the models of Sun et al. \cite{Bfield2} and Pshirkov et al. \cite{Bfield3}, the exclusion regions have been tightened and loosened, respectively (see Appendix \ref{sec:uncertainties}).  We thus conclude that  the constraints reported in Figure \ref{fig:cpr} are reasonable and representative.
In addition to the uncertainty related to the choice of magnetic field model, some other systematic uncertainties have also been discussed in Appendix \ref{sec:uncertainties}.
Among the discussed systematic effects, the flux measurement uncertainty has relatively large effect on the final constraints.
But all the limits presented in Figure \ref{fig:cpr} are taken at face value \cite{Meyer13,hess13pks2155,fermi16alp,zc16alp}, so the influence by such an uncertainty is ignored in the comparisons of Figure \ref{fig:cpr}.

Our results demonstrate that the very high energy gamma-ray observations of the bright Galactic sources also play important roles in probing ALP properties.
The upcoming next generation Cherenkov telescope, CTA \cite{cta}, has significantly enhanced detection sensitivity in the energy range from 50 GeV to 100 TeV, which would remarkably advance the ALP searches. Here we present the projected exclusion region of the CTA based on Monte-Carlo simulations (see Appendix \ref{sec:sim} for details of the simulation).
We find that 50-hour CTA observations of the 10 sources studied in this work will constrain ALP parameters in a much wider region (see gray region in Figure \ref{fig:cpr}).
It will probe the ALPs with $10^{-7}~{\rm eV}<m_{\rm a}<10^{-6}~{\rm eV}$ and $\theta_1{\cal N}\leq 3$, where ALPs could make up $1/9$ of the DM content of the Universe.
It is also reasonable to speculate that the CTA observation of some extragalactic sources, such as PKS 2155-304, would notably improve the current limits around the ALP masses of $m_{\rm}\sim10^{-8}-10^{-7}\,{\rm eV}$.

\section{summary}

ALP is one promising type of cold dark matter candidate, which can also solve the issue of the TeV transparency of the Universe.
In this work, we constrain the ALP parameters with \hess observations of bright TeV sources within the Milky Way.
Through our analysis, we suggest that the Galactic TeV $\gamma$-ray sources, which are usually observed by ground based Cherenkov telescope, can provide insights into the ALP physics as well.
Specifically, by analyzing the H.E.S.S. observations of 10 Galactic sources, the high-mass part of the parameter space of explaining the TeV transparency is constrained.
Moreover, Monte-Carlo simulations show that the next generation Cherenkov telescope, CTA, can probe a wider region of the parameter space.
We therefore expect that the ALP interpretation of the low opacity of the universe will be unambiguously tested in the near future.

Note that some on-orbit/future space-borne instruments, such as the Dark Matter Particle Explorer (DAMPE) \cite{DAMPE1,DAMPE2} and High Energy cosmic-Radiation
Detection Facility (HERD) \cite{zhang14herd}, are also sensitive to the photons in TeV energy range. Though with much smaller effective area, they have significantly higher energy resolution, thus may also be able to help us better understand the ALP properties.
Together with the upcoming ALP-II \cite{ALPSII2013} and IAXO \cite{IAXO2015} experiments, significant progresses on revealing the ALP properties are expected in the next decade.

\begin{acknowledgments}
All the spectra used in this paper are from the H.E.S.S. website.
This research has made use of the CTA instrument response functions provided by the CTA Consortium and Observatory, see http://www.cta-observatory.org/science/cta-performance/ (version prod3b-v1) for more details.
This work is supported by the National Key Research and Development Program
of China (No. 2016YFA0400200), the National Natural Science Foundation
of China (Nos. 11525313, 11722328, 11773075, U1738210, U1738136), the 100 Talents Program of
Chinese Academy of Sciences, and the Youth Innovation Promotion Association
of Chinese Academy of Sciences (No. 2016288).
\end{acknowledgments}

\bibliographystyle{apsrev4-1-lyf}
\bibliography{alp_hess}

\widetext
\appendix

\section{Null distribution and 95\% confidence level}
\label{sec:threshold}
Statistically, the 95\% confidence level (C.L.) exclusion region corresponds to the parameters leading to an increase of the $\chi^2$ by $>5.99$ comparing to the best fit.
However, due to the non-linear dependence of the spectral irregularities on the ALP parameters and possible systematics in the observations, this threshold value $\lambda_{\rm thr}$ may be biased. In this work we adopt the $\lambda_{\rm thr}$ derived from Monte Carlo simulations to set the 95\% C.L. limits.

The $\lambda$ values are given by the $\Delta\chi^2$ between the best fit with a certain ALP mass and coupling (i.e., only fit the nuisance parameters) and that with them free to vary. Thus the $\lambda_{\rm thr}$ values are different for different sets of ALP parameters ($m_a$, $g_{a\gamma}$) and we need to simulate the distribution of $\lambda(m_{a},g_{{a}\gamma})$ (alternative distribution) for the complete parameter space, which is however computational expensive.
Following Ref. \cite{fermi16alp}, in this work we assume that the probability distribution of the alternative hypothesis can be approximated with the null distribution. 
It is found that such an assumption would yield conservative limits \cite{fermi16alp}.

\begin{figure}
\centering
\includegraphics[width=0.45\textwidth]{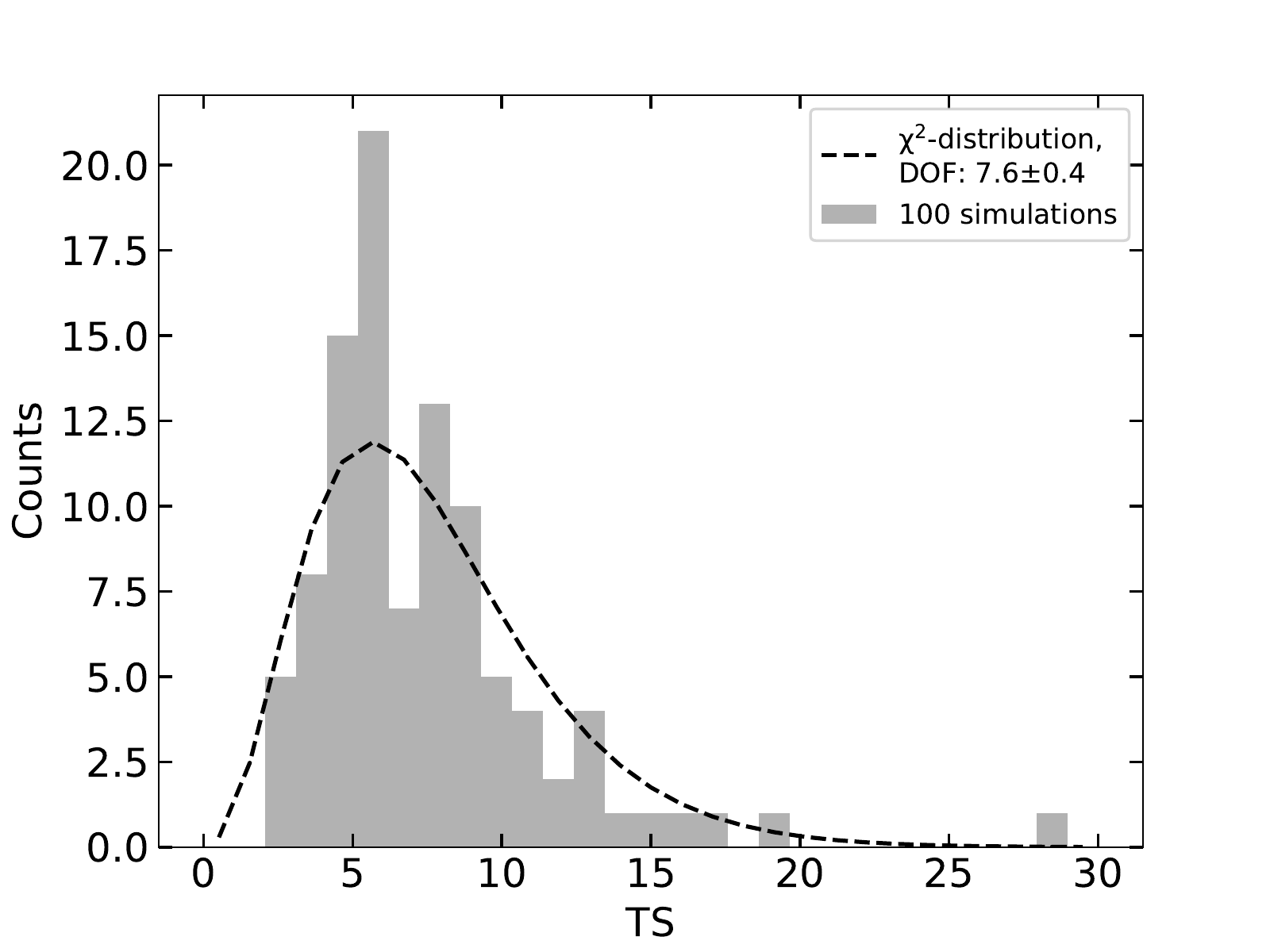}
\caption{The null distribution of the TS values found in 100 Monte Carlo simulations. The distribution follows a $\chi^2$ distribution with a degree of freedom $d=7.6\pm0.4$. 
}
\label{fig:tsdistri}
\end{figure}

Accurate simulations of the observed spectra of the 10 sources require the \hess exposures to these sources, which are however unknown.
We therefore generate the simulated spectra in the following way. For each source, the simulated spectrum is set to have the same energy bins as the observed one. In each energy bin, the nominal value of the flux is randomly generated based on the Gaussian distribution with its mean being the model expected flux calculated by the best fit null model and the sigma being the uncertainty of the observation. {\it The error bar of the flux in the simulated spectrum is set to that of the observed one.} With this method, we generate 100 sets of Monte Carlo null spectra of the 10 sources. For each set of the simulated spectra, we perform the same analysis as used in the main article to derive the best fit $\hat{\chi}^2$ for both null model and ALP model.
The distribution of the $TS=\hat{\chi}^2_{\rm null}-\hat{\chi}^2_{\rm wALP}$ for 100 simulations is shown in Figure \ref{fig:tsdistri}. A non-central $\chi^2$ fit to the null distribution results in degree of freedom (DOF) $d=7.6\pm0.4$ and non-centrality $s\approx0$ (indicating that a standard $\chi^2$ function can fit the distribution well).
From the best fit distribution, we derive the threshold $\lambda_{\rm thr}=14.99\pm0.53$ above which the ALP parameters could be constrained at 95\% confidence level. The $1\sigma$ error bar is due to the limited number of our simulations and is derived from bootstrapping the null distribution $10^4$ times.
For conservativeness, the upper bound $\lambda_{\rm thr}=15.52$ is used to set the limits.

\section{Related Uncertainties}
\label{sec:uncertainties}
Our analysis suffers from some systematic uncertainties, we will discuss them here.

\begin{figure}
\centering
\includegraphics[width=0.45\textwidth]{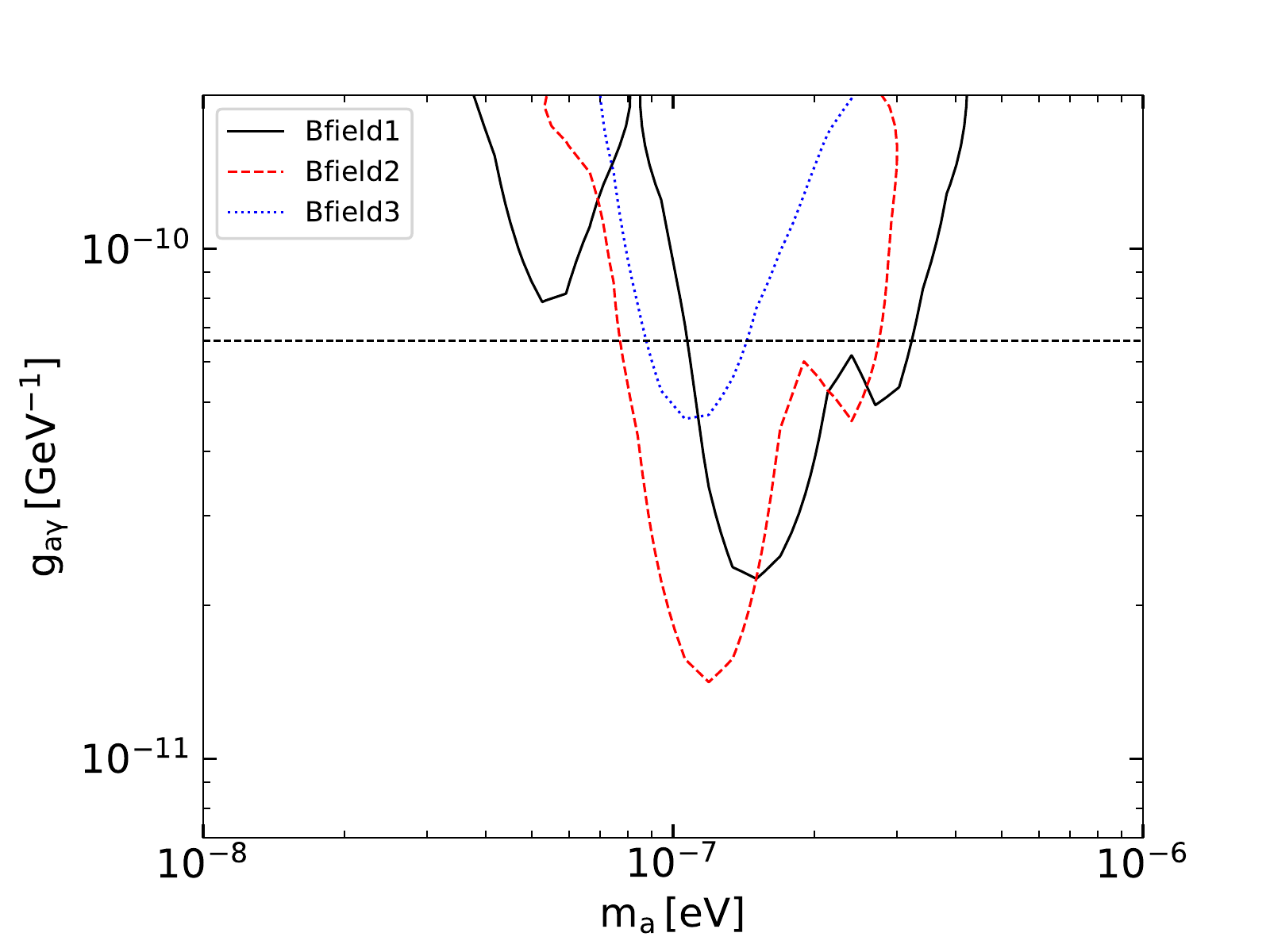}
\caption{The 95\% C.L. exclusion regions of ALP mass $m_{a}$ and photon-ALP coupling constant $g_{a\gamma}$ obtained in the combined analysis. The three lines are for different Milky Way magnetic field models. Also plotted is the upper limit of the photon-ALP coupling set by CAST \cite{CAST2017} experiment (horizontal line).}
\label{fig:combined}
\end{figure}

{\tt Galactic magnetic field}: For the regular component of the Galactic magnetic field, in the analysis of the main text, we adopt the model of Jansson \& Farrar \cite{Bfield1} (Bfield1). To test the uncertainty related to the choice of magnetic model, we also take into account two additional models developed by Sun et al. \cite{Bfield2} and Pshirkov et al. \cite{Bfield3}, which are denoted as Bfield2 and Bfield3, respectively.
As is shown in Figure \ref{fig:combined}, adopting the model of Bfield2 gives the most stringent constraints on the ALP parameters, while the weakest one is obtained in the model of Bfield3. Comparing to the model of Bfield1, the area of the exclusion region below CAST limit changes by +57\% and -80\% for Bfield2 and Bfield3, respectively. Although the improvement is marginal for the result of Bfield3, it restricts the allowed ALP parameter space by an independent method, thus is complementary to the CAST constraints.

\begin{figure}
\centering
\includegraphics[width=0.45\textwidth]{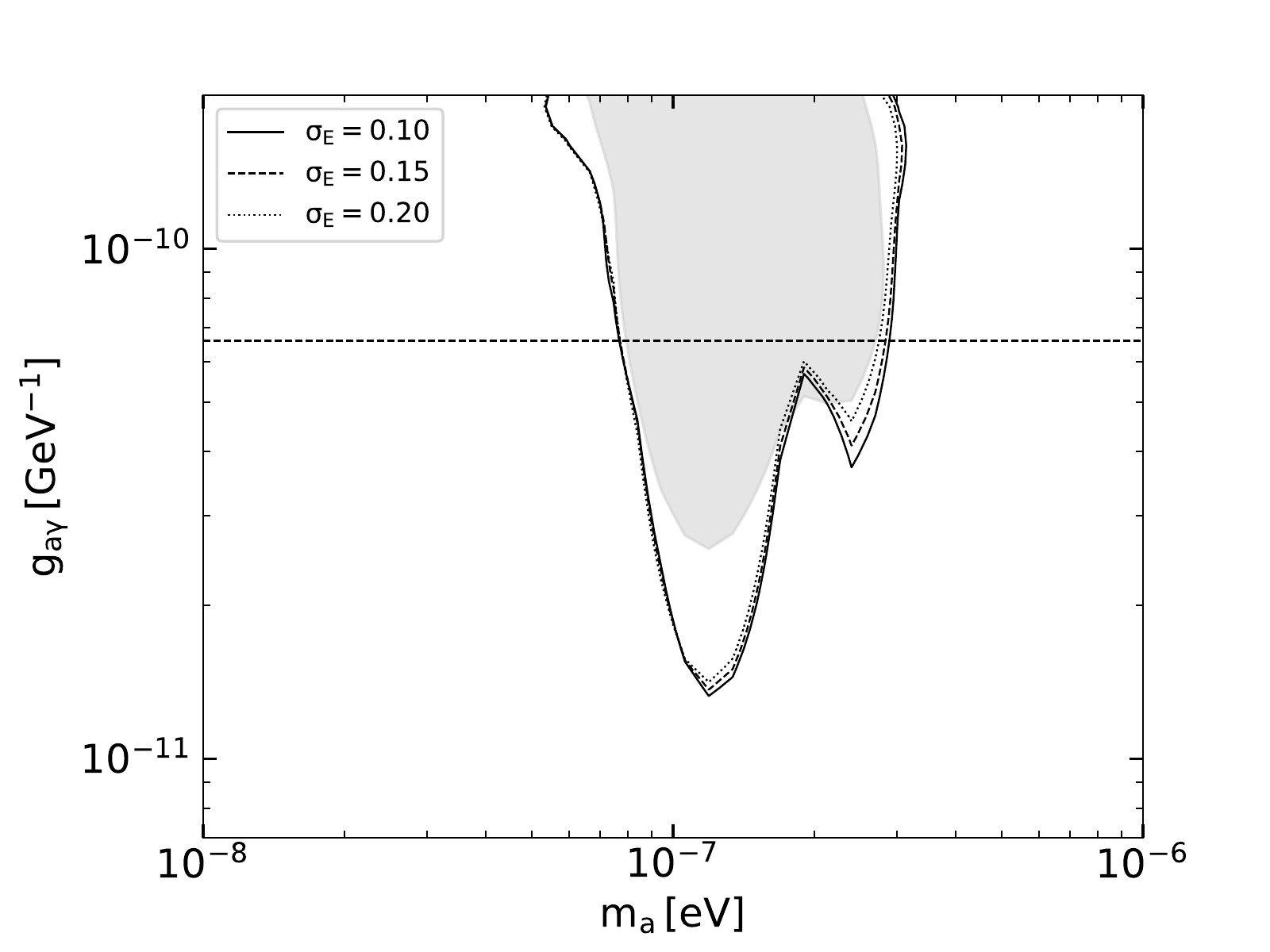}
\caption{The 95\% C.L. exclusion regions of ALP parameters assuming three different energy resolution of \hess The shaded area further takes into account a 20\% uncertainty on flux measurements for $\sigma_E=0.2$. All these results are for Bfield2.}
\label{fig:unc}
\end{figure}

{\tt Energy resolution}: In this work, we have assumed a single value of 20\% of the detector's energy resolution ($\sigma_E$). In fact, the energy resolution of the \hess is dependent of the observation mode and the source incidence angle and varies with the photon's energy. Inaccurate use of the energy resolution may cause systematics to the results. The energy resolution of \hess reported in the literature range from $\sim$10\% to $\sim$20\% \cite{hess13pks2155,hess_line1,hess_line2}. We therefore test our results for three values of the energy resolution, 10\%, 15\% and 20\%, to show how the final constraints would be affected. Figure \ref{fig:unc} shows the results.
The modification of the energy resolution between 10\% and 20\% has only marginal effects on the constraints and our fiducial result with 20\% $\sigma_E$ is the most conservative one.

{\tt Energy binning}: The spectra used by us are collected from the literature, we could not examine the systematics associated with the energy binning. As is shown in Ref. \cite{hess13pks2155}, which also focused on studying ALP signals with \hess data, varying the binning does lead to a certain level of fluctuations on the irregularity caused by the ALP modulation. However, the method used in their analysis, which measures the level of irregularity in the spectrum of extragalactic source, is not the same as that in this work. In our previous work of adopting also $\chi^2$ fit to analyze the Fermi-LAT data of the Galactic sources, we find that the energy binning has only a minor influence on the results \cite{ic443alp}.

{\tt Flux measurements}:  A number of instrument effects (e.g. effective area) may generate systematic uncertainty on flux measurements. In this work, we use a simple way to estimate how such a uncertainty would influence the final constraints. We simply assume the total systematic uncertainty on the flux measurements is 20\% \cite{hess06crab} and add it in quadrature to the statistical errors. With this additional flux uncertainty included, we get somewhat weaker constraints on the ALP parameters (see shaded area in Figure \ref{fig:unc}). 

\section{CTA perspective}
\label{sec:sim}
If no ALP signal is found in the future CTA observations of these bright Galactic gamma-ray sources, the allowed region for the ALP model, especially those can account for the low opacity of the Universe, will be further constrained. Here we perform Monte Carlo simulations to derive the projected exclusion regions that CTA can provides.

\begin{figure}
\centering
\includegraphics[width=0.45\textwidth]{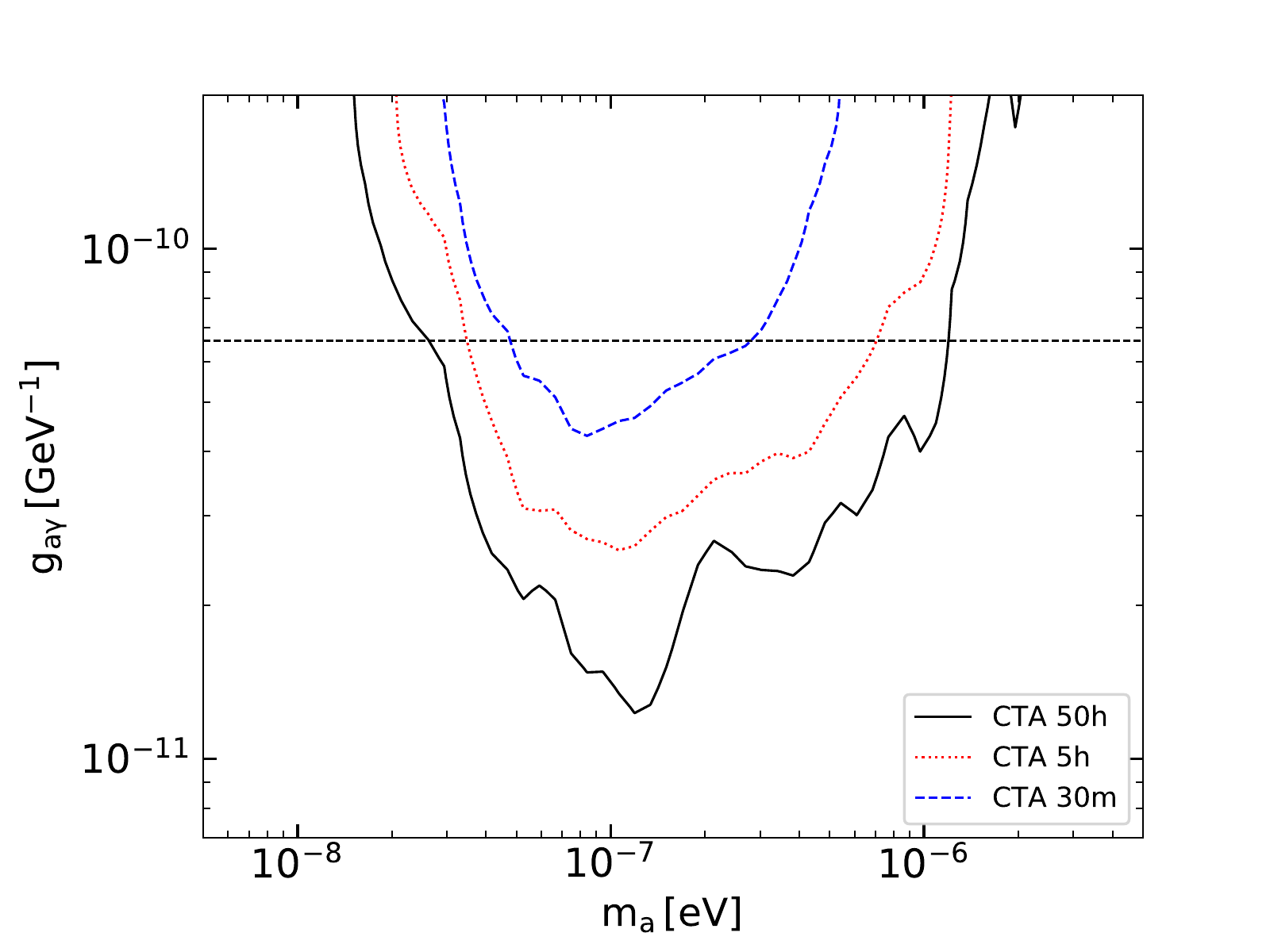}
\caption{The prospective exclusion regions of ALP parameters by future CTA experiment \cite{cta} based on its 30-minute, 5-hour and 50-hour observations.}
\label{fig:cta}
\end{figure}

We generate pseudo photons according to Poisson statistics and the best-fit null model obtained in the above analyses of H.E.S.S. spectra. The energy-dependent collection area of CTA instruments is considered when deriving the CTA exposure to each source. Since all these 10 sources are in the south sky, we utilize the instrument response functions (IRFs) for the southern sites of the CTA Observatory \footnote{\url{https://www.cta-observatory.org/science/cta-performance/}}. To be conservative, we adopt the IRFs of 40$^\circ$ zenith angle. The generated pseudo photons are then binned into 30 logarithmically spaced energy bins from 0.1 TeV to 50 TeV. We do not take into account the data outside this energy range where CTA is also sensitive to avoid extrapolation of the spectra. Similar to the analysis of real H.E.S.S. observations, we ignore the bins with $\tilde{\phi}_i<3\delta_i$. The same $\chi^2$ analysis procedure is used to fit the simulated spectra to derive the combined $\Delta\chi^2$ map for each simulation. The energy dispersion function is approximated by a Gaussian with its $\sigma$ being the energy-dependent energy resolution of the CTA.  Totally, 100 simulations are performed. The mean exclusion regions derived from Monte-Carlo simulations for 30-minute, 5-hour and 50-hour CTA observations are plotted in Figure \ref{fig:cta}.

\section{Individual searching results}
\label{sec:9srcs}
The $\Delta\chi^2$ maps as a function of ALP mass $m_{a}$ and photon-ALP coupling constant $g_{a\gamma}$ for other 9 sources among our sample (except HESS J1640-465, which has been presented in the main text) are plotted here.

\begin{figure*}[!hb]
\centering
\includegraphics[width=0.33\textwidth]{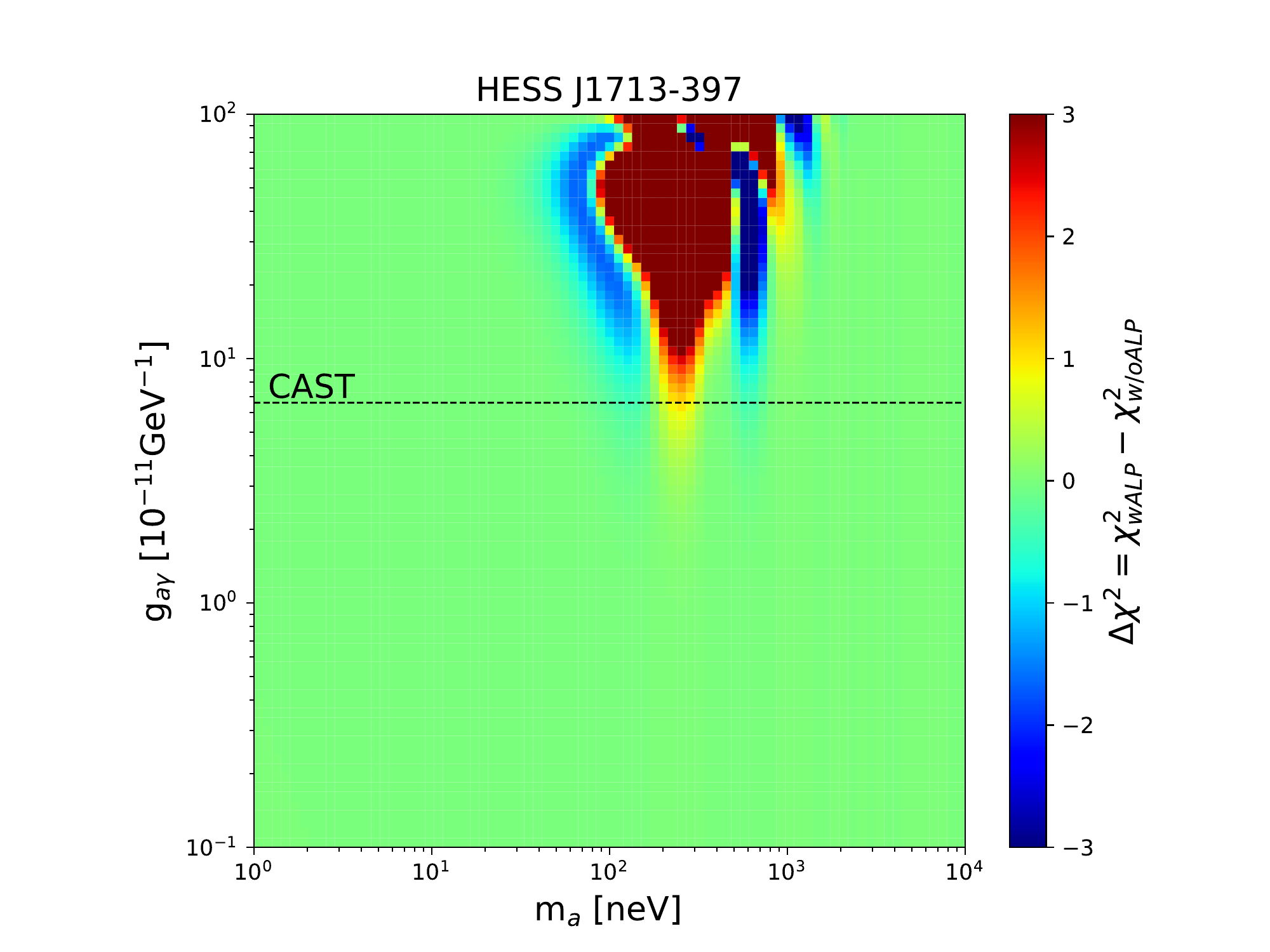}
\includegraphics[width=0.33\textwidth]{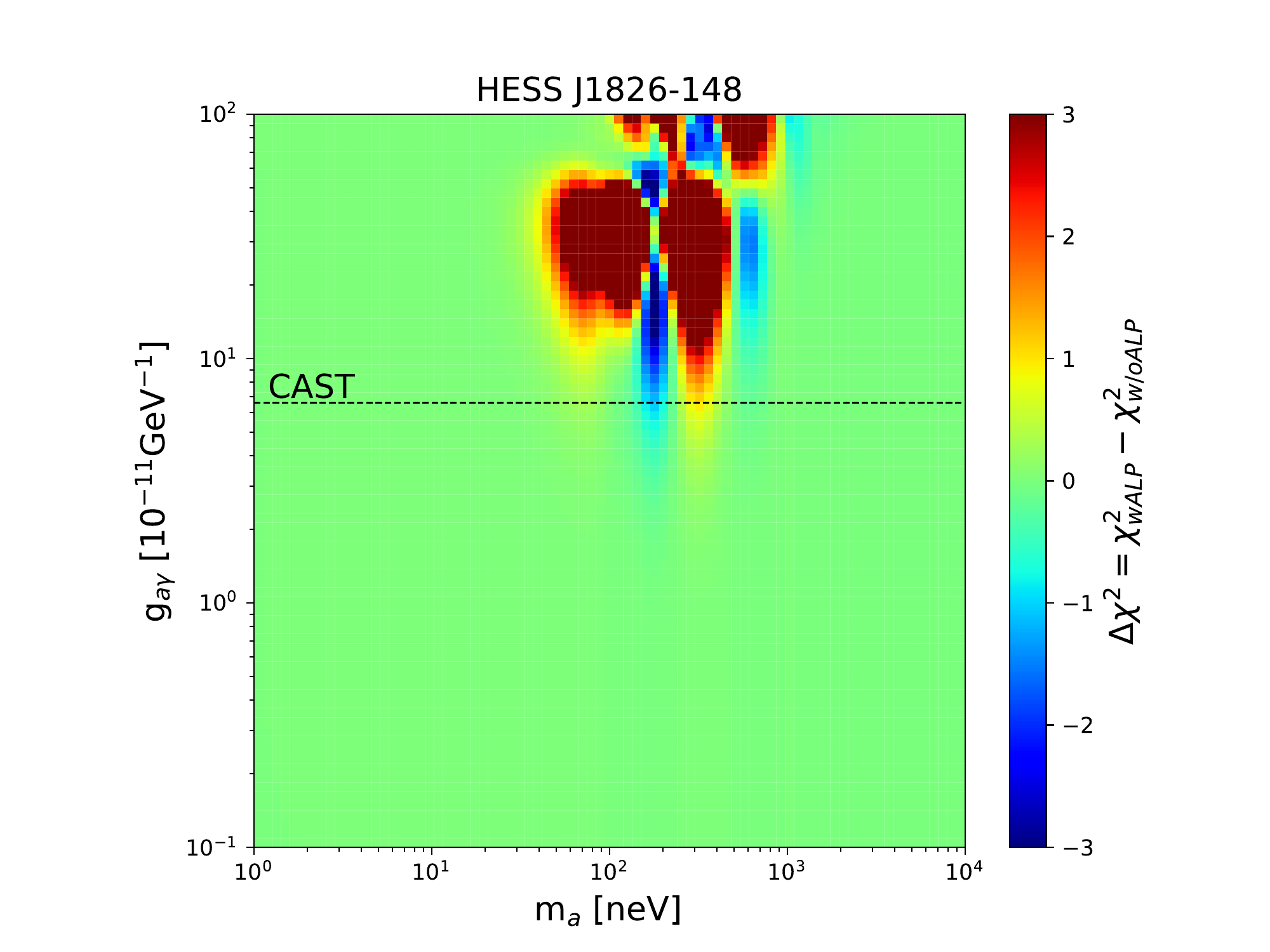}
\includegraphics[width=0.33\textwidth]{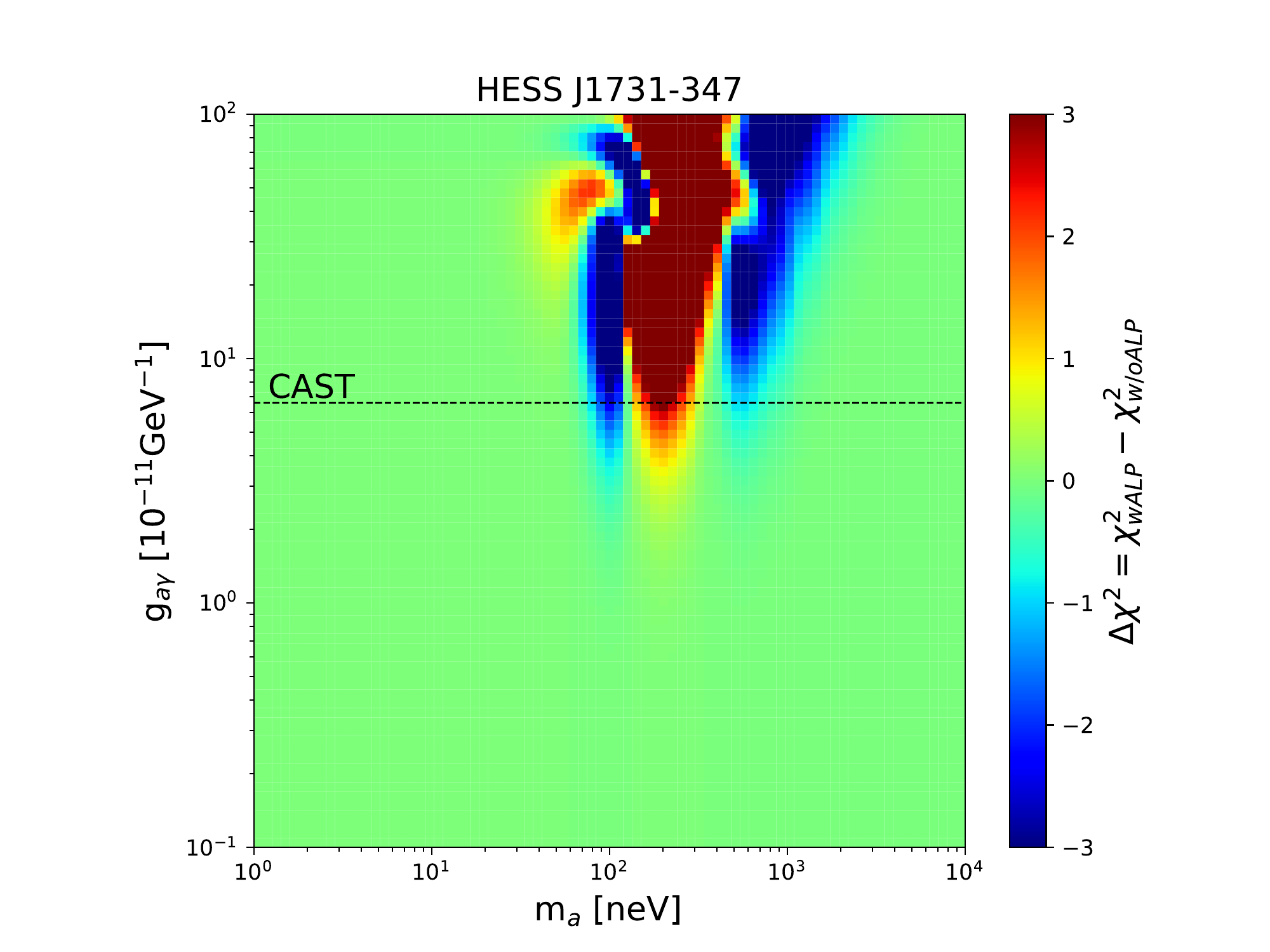}\\
\includegraphics[width=0.33\textwidth]{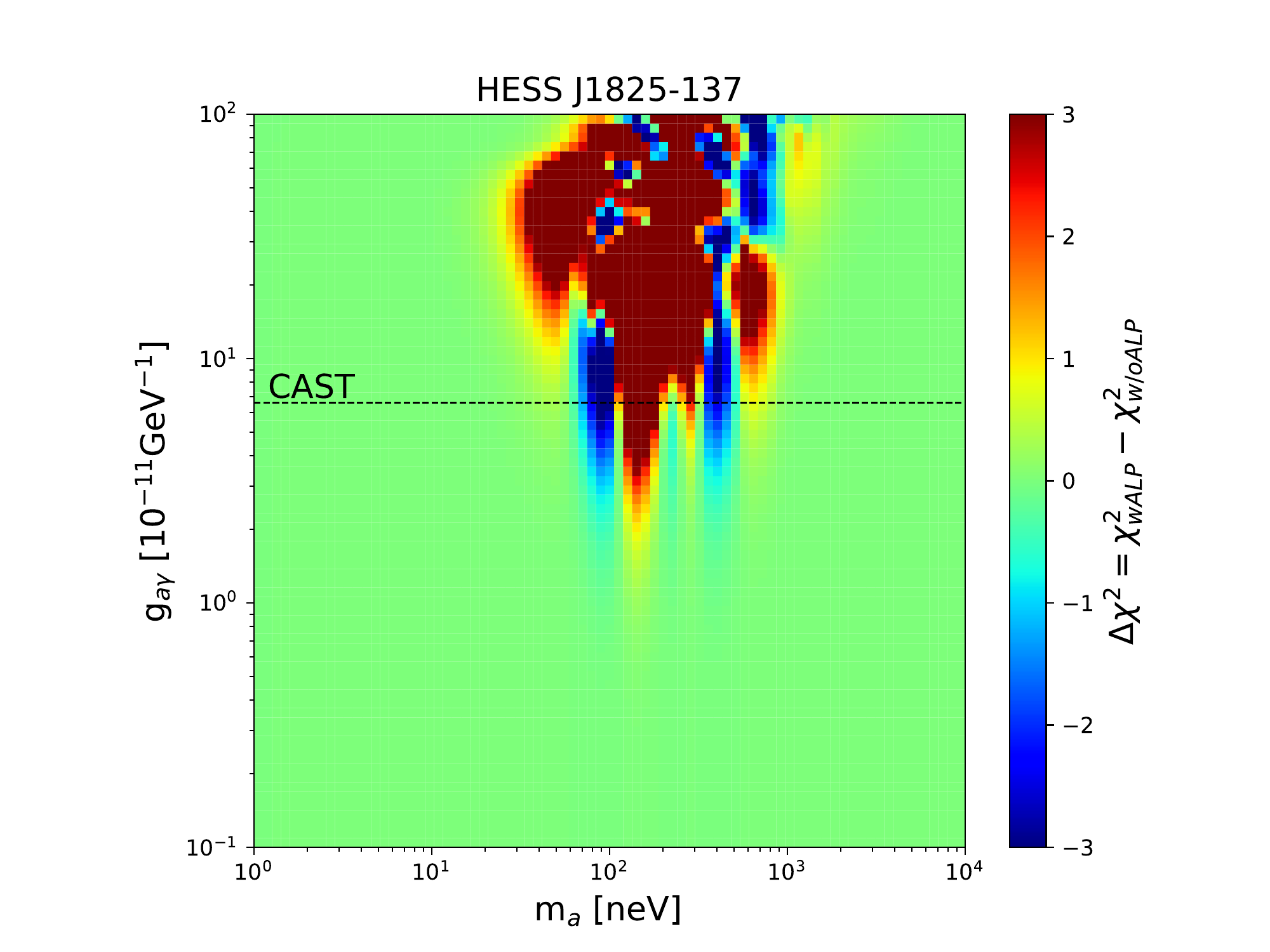}
\includegraphics[width=0.33\textwidth]{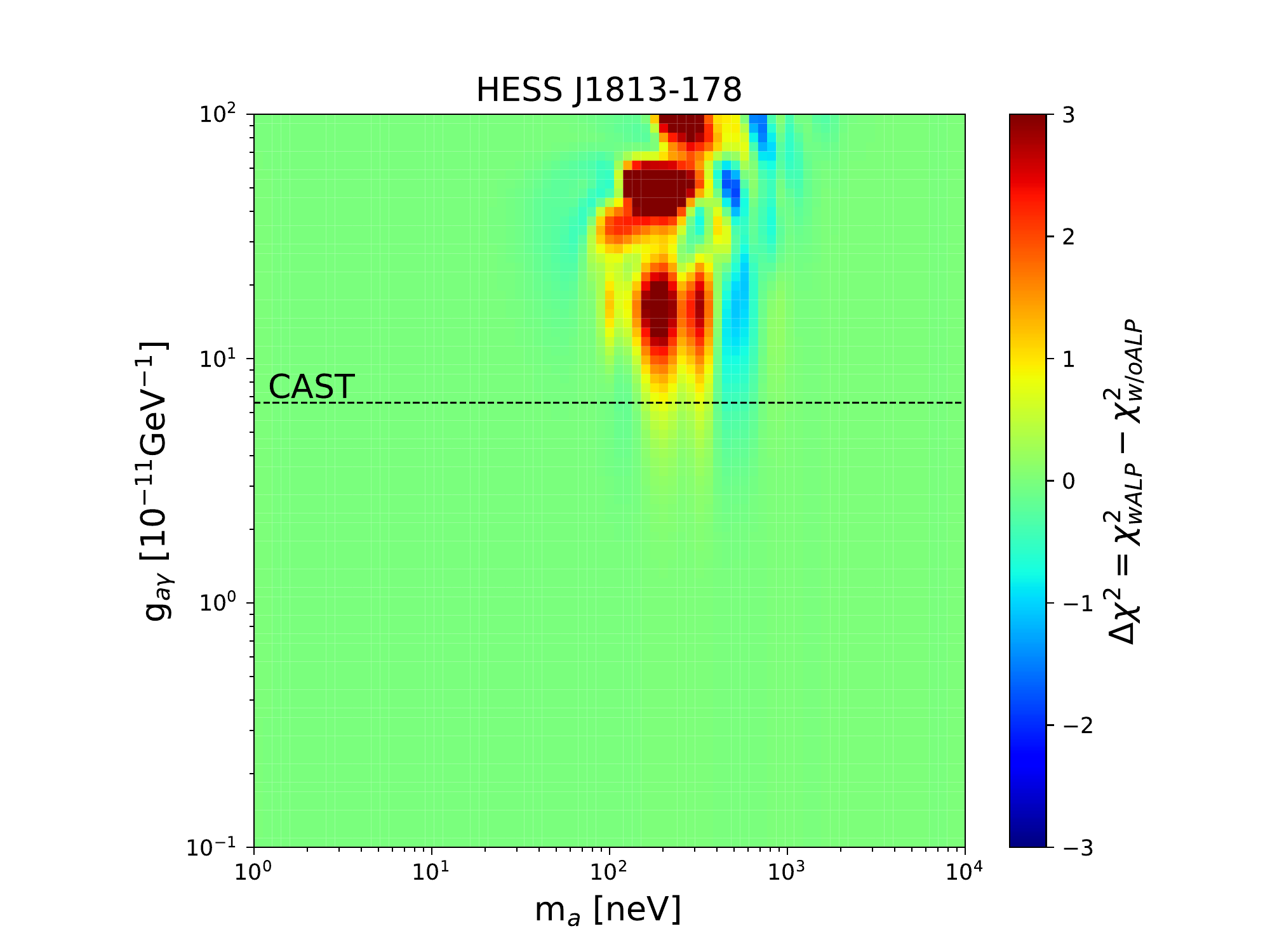}
\includegraphics[width=0.33\textwidth]{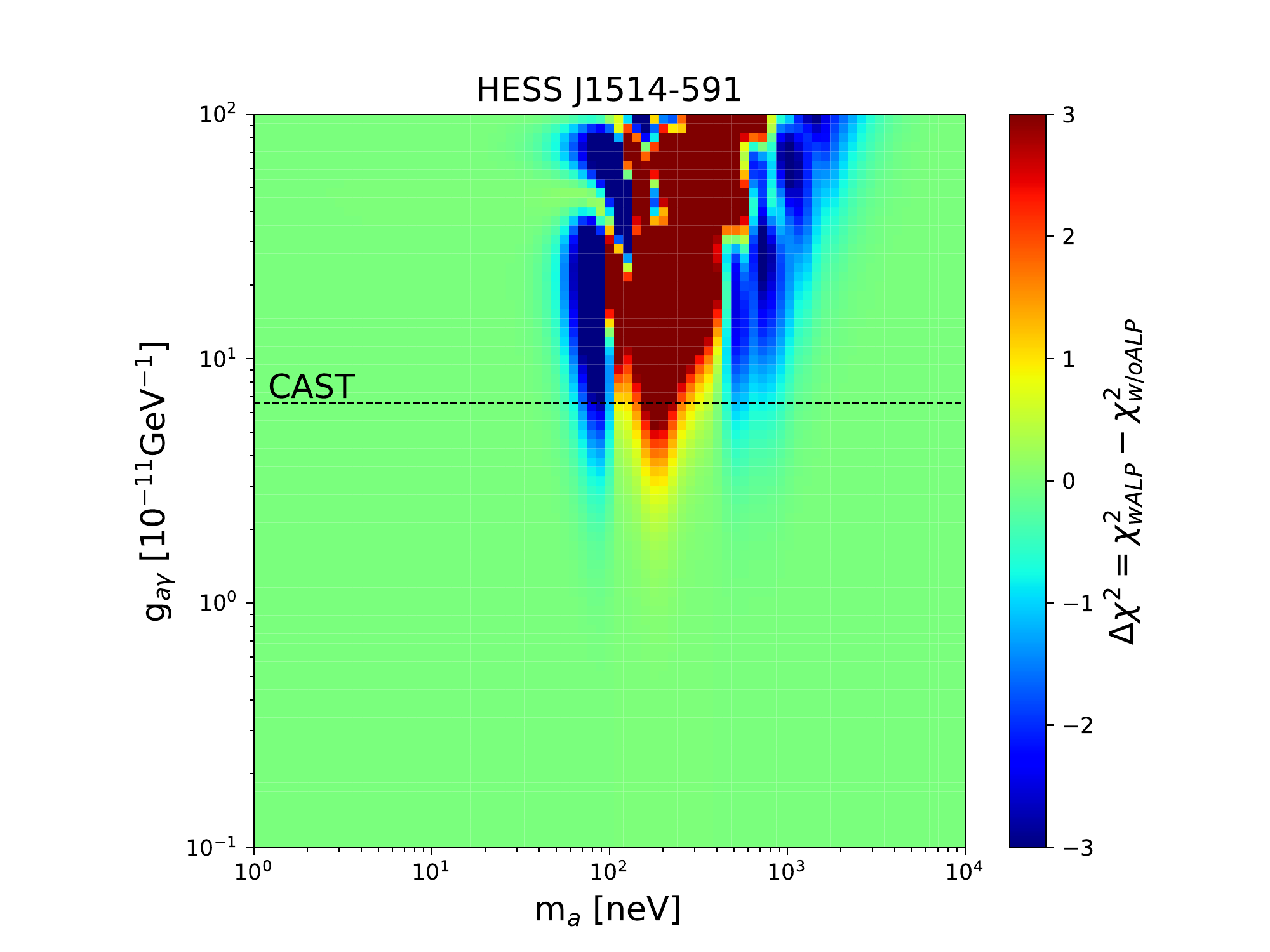}\\
\includegraphics[width=0.33\textwidth]{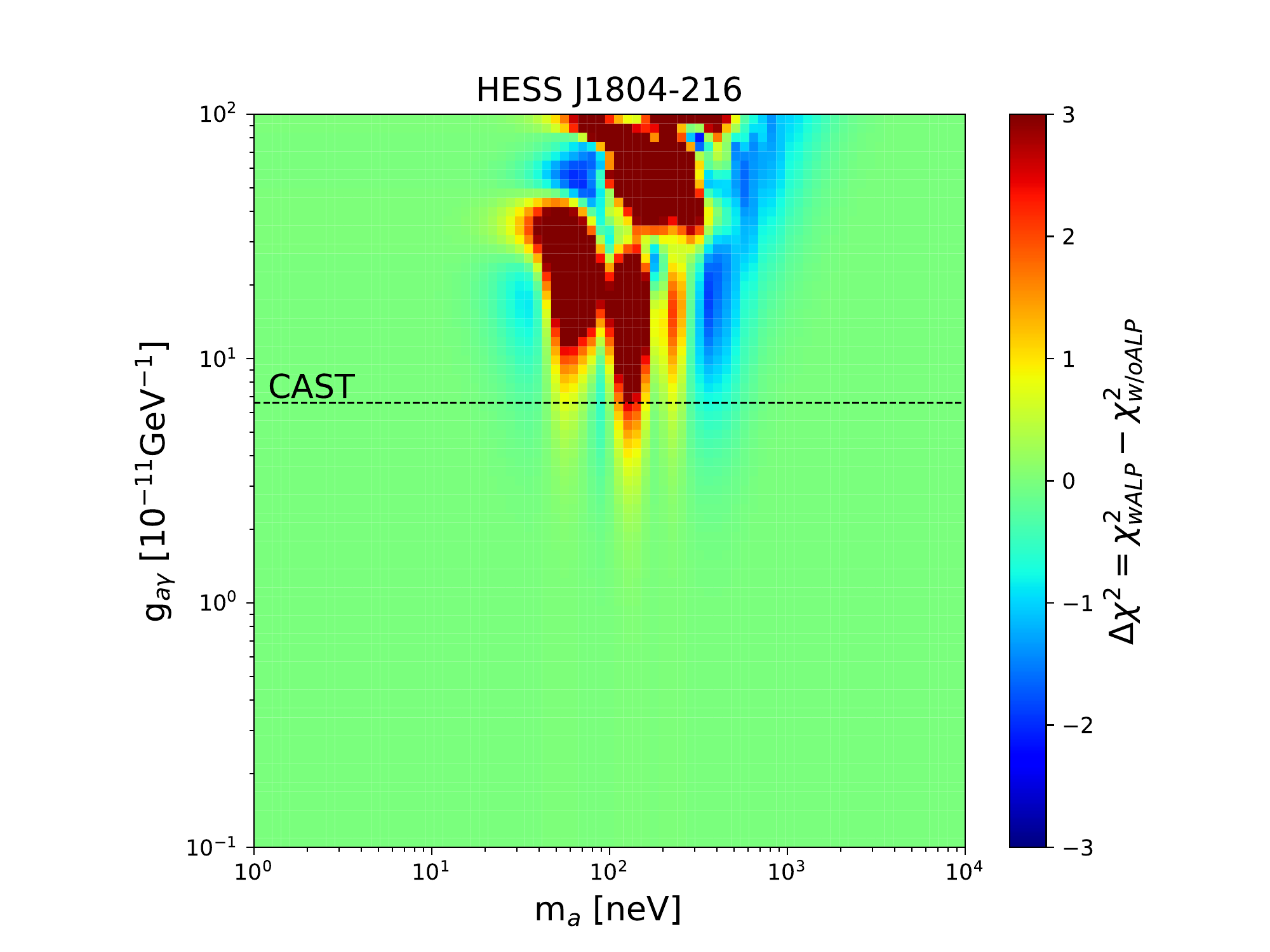}
\includegraphics[width=0.33\textwidth]{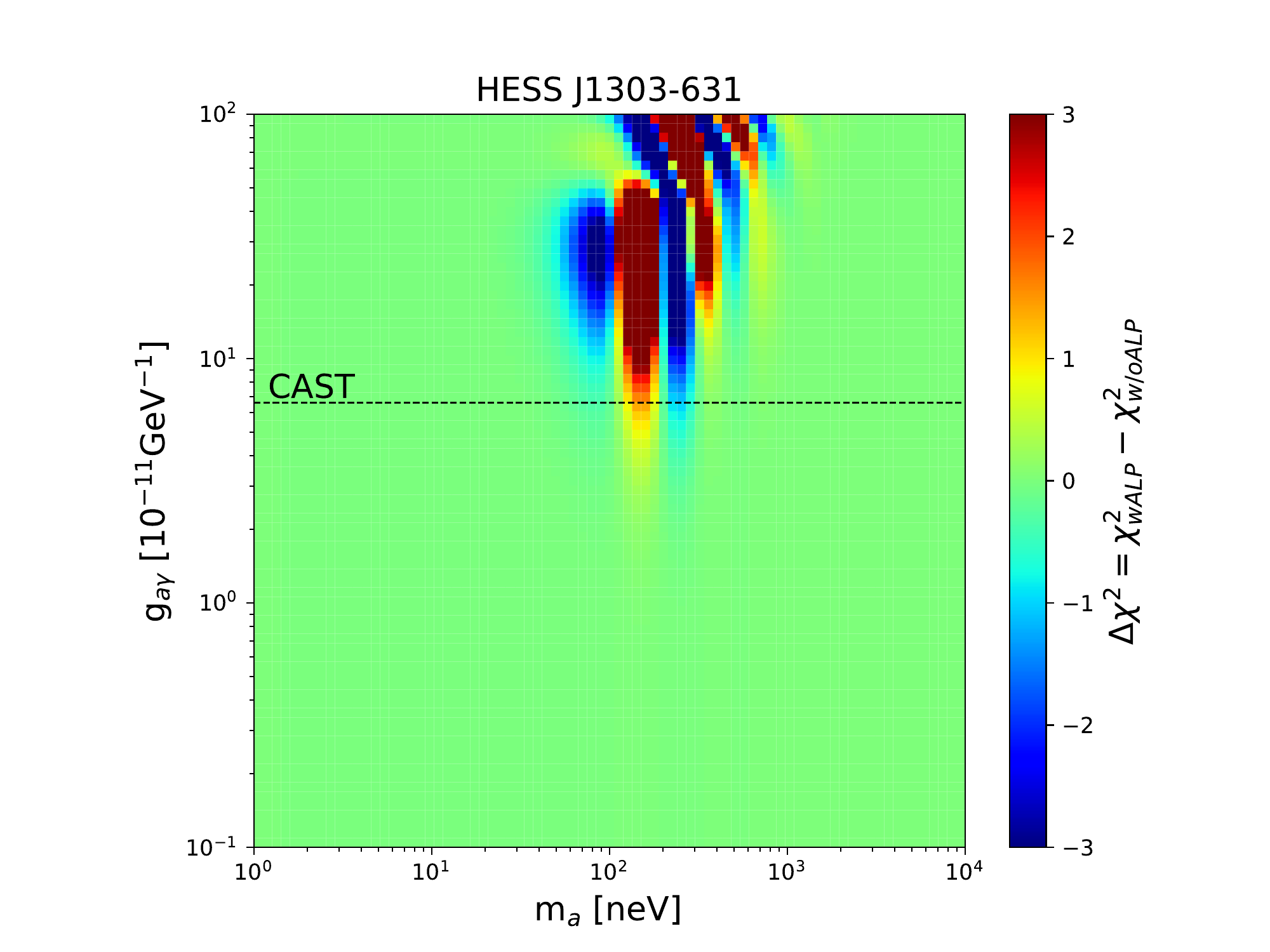}
\includegraphics[width=0.33\textwidth]{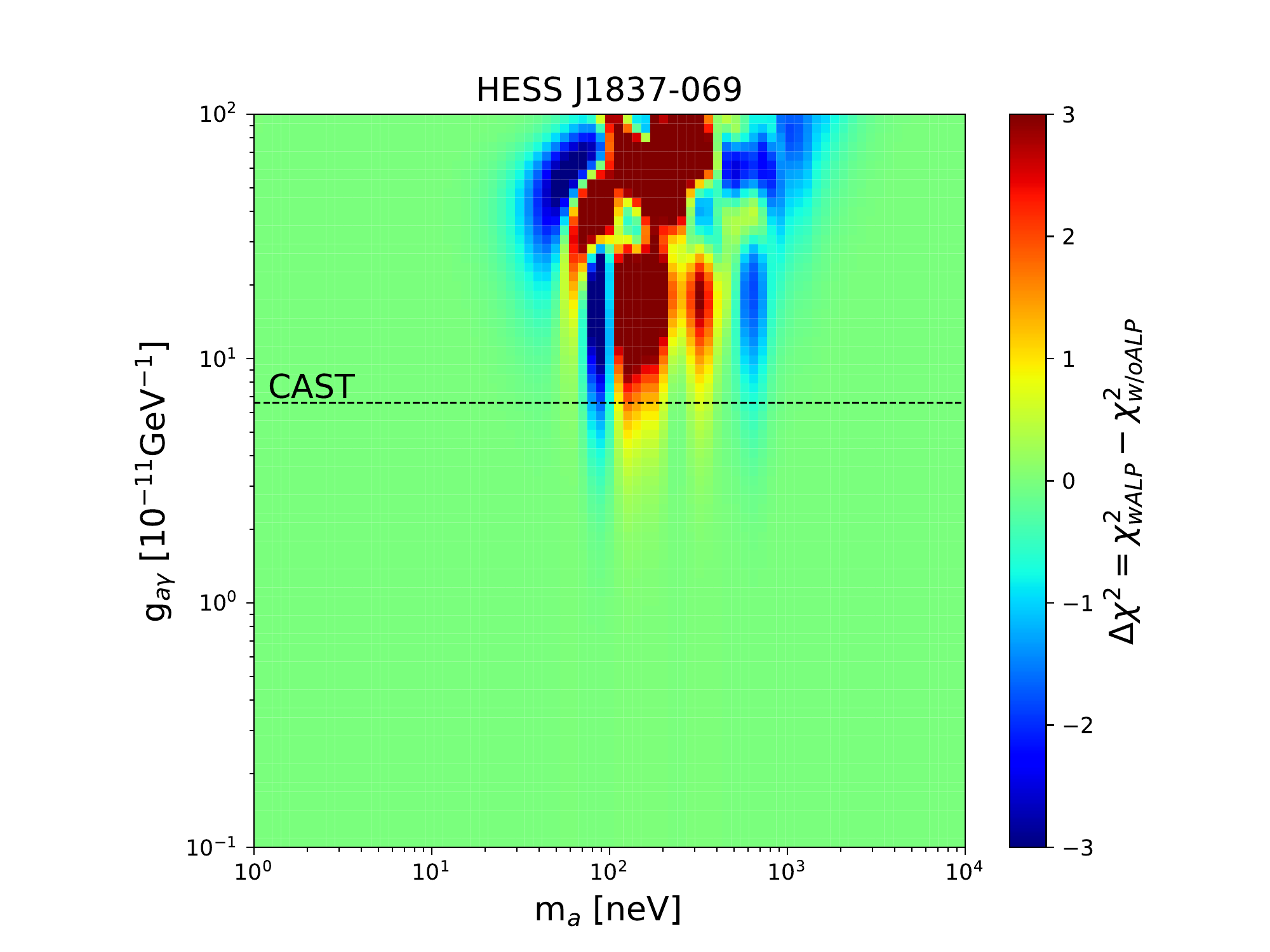}\\
\caption{The maps of $\Delta\chi^2$ as a function of ALP mass $m_{a}$ and photon-ALP coupling constant $g_{a\gamma}$ for 9 sources among our sample. The upper limit of $g_{a\gamma}$ set by CAST \cite{CAST2017} experiment is also plotted as a reference (dashed horizontal line).}
\label{fig:3srcs}
\end{figure*}

\end{document}